# The statistical properties of galaxy morphological types in compact groups of Main galaxies from the SDSS Data Release 4

Xin-Fa Deng    Peng Jiang    Cong-Gen He    Cheng-hong Luo    Ji-Zhou He

School of Science, Nanchang University, Jiangxi, China, 330047

**Abstract**    In order to explore the statistical properties of galaxy morphological types in compact groups (CGs), we construct a random group sample which has the same distributions of redshift and number of member galaxies as those of the CG sample. It turns out that the proportion of early-type galaxies in different redshift bins for the CG sample is statistically higher than that for random group sample, and with growing redshift z this kind of difference becomes more significant. This may be due to the existence of interactions and mergers within a significant fraction of SDSS CGs. We also compare statistical results of CGs with those of more compact groups and pairs, but do not observe as large statistical difference as Hickson (1982)' results.

**Keywords**    Galaxy: fundamental parameters-- galaxy: large scale structure

## 1. Introduction

Compact groups of galaxies (CGs), defined by their small number of members (< 10) and their compactness, are a group of special and interesting objects in the universe for studying the properties of galaxies and understanding of the overall structure of the universe. Many studies show that the proportion of different morphological type galaxies in compact groups is significantly different from that in the field (Hickson 1982, Williams & Rood 1987, Sulentic 1987, Hickson et al. 1988, Rood & Williams 1989, Prandoni et al. 1994, Lee et al. 2004). This is of particular importance to studies of the physical nature of compact groups(see the review by Hickson 1997).

By analyzing the morphological types of 451 member galaxies in the catalog groups, Hickson (1982) found that the groups contain fewer spiral galaxies than a comparable sample of field galaxies, and the proportion of spiral galaxies decreases from 60% in the least compact groups to 20% in the most compact. Using the Hickson Compact Group (HCG) catalog, Hickson et al. (1988) calculated the proportion $P_{late}$ of late type galaxies in the HCGs ($P_{late}$ = 0.49). In the Southern Compact Group (SCG) catalog, Prandoni et al. (1994) obtained $P_{late}$ = 0.59 for the SCGs. Both values are substantially lower than those found for field galaxy samples ($P_{late}$ =0.82, Gisler 1980, Nilson 1973). Similarly, Rood & Williams (1989) found that compact groups contain a significantly smaller fraction of late-type (spiral and irregular) galaxies than do their neighborhoods.

Using 175 CGs identified from the Sloan Digital Sky Survey (SDSS), Lee et al. (2004) explored morphology-environment effects in SDSS CGs. They found that the rest-frame colors of CG galaxies indeed differ from those of field galaxies — at least for $M_{u^*} - M_{g^*}$, $M_{g^*} - M_{r^*}$, and even $M_{r^*} - M_{i^*}$, and concluded that SDSSCGs contain a relatively higher fraction of Elliptical galaxies than does the field. N-body simulations pioneered by Toomre (1977) indicated that the end-product of merging Spirals can be an Elliptical galaxy. So, above results showed that there is strong evidence of interactions and mergers within a significant fraction of SDSS CGs.

In this paper, we use a new catalog of CGs, and further explore the statistical properties of morphological types of member galaxies in CGs. Our paper is organized as follows. In section 2, we describe the galaxy data to be used. The group identification algorithm and the CG catalog are discussed in section 3. In section 4, we analyse the statistical properties of morphological types of member galaxies in CGs. Our main results and conclusions are summarized in section 5.

## 2. Galaxy data

The Sloan Digital Sky Survey (SDSS) is one of the largest astronomical surveys to date. The completed survey will cover approximately 10000 square degrees. York et al. (2000) provided the technical summary of the SDSS. The SDSS observes galaxies in five photometric bands (u, g, r, i, z) centered at (3540, 4770, 6230, 7630, 9130 Å). The imaging camera was described by Gunn et al. (1998), while the photometric system and the photometric calibration of the SDSS imaging data were roughly described by Fukugita et al. (1996), Hogg et al. (2001) and Smith et al. (2002) respectively. Pier et al. (2003) described the methods and algorithms involved in the astrometric calibration of the survey, and present a detailed analysis of the accuracy achieved. Many of the survey properties were discussed in detail in the Early Data Release paper (Stoughton et al. 2002). Galaxy spectroscopic target selection can be implemented by two algorithms. The MAIN Galaxy sample (Strauss et al. 2002) targets galaxies brighter than $r_p < 17.77$ (r-band apparent Petrosian magnitude). Most galaxies of this sample are within redshift region: $0.02 \leq z \leq 0.2$. The Luminous Red Galaxy (LRG) algorithm (Eisenstein et al. 2001) selects galaxies to $r_p < 19.5$ that are likely to be luminous early-types, using color-magnitude cuts in g, r, and i. Because most LRGs are within redshift region: $0.2 \leq z \leq 0.4$, two samples mentioned above actually represent the distribution of galaxies located at different depth.

The SDSS has adopted a modified form of the Petrosian (1976) system for galaxy photometry, which is designed to measure a constant fraction of the total light independent of the surface-brightness limit. The Petrosian radius $r_p$ is defined to be the radius where the local surface-brightness averaged in an annulus equals 20 percent of the mean surface-brightness interior to this annulus, i.e.

$$\frac{\int_{0.8r_p}^{1.25r_p} dr 2\pi \, rI(r)/[\pi(1.25^2 - 0.8^2)r^2]}{\int_0^{r_p} dr 2\pi \, rI(r)/[\pi \, r^2]} = 0.2$$

where I(r) is the azimuthally averaged surface-brightness profile. The Petrosian flux $F_p$ is then defined as the total flux within a radius of $2r_p$, $F_p = \int_0^{2r_p} 2\pi \, rdrI(r)$. With this definition, the Petrosian flux (magnitude) is about 98 percent of the total flux for an exponential profile and about 80 percent for a de Vaucouleurs profile. The other two Petrosian radii listed in the Photo output, $R_{50}$ and $R_{90}$, are the radii enclosing 50 percent and 90 percent of the Petrosian flux, respectively.

The SDSS4 sky coverage can be separated into three regions. Two of them are located in the north of the Galactic plane, one region at the celestial equator and another at high declination. The third lies in the south of the Galactic plane, a set of three stripes near the equator. Each of these regions covers a wide range of survey longitude.

In our work, we consider the Main galaxy sample. The data is download from the Catalog Archive Server of SDSS Data Release 4 (Adelman-McCarthy et al. 2006) by the SDSS SQL Search (with SDSS flag: bestPrimtarget=64) with high-confidence redshifts (Zwarning $\neq 16$ and Zstatus $\neq 0$, 1 and redshift confidence level: zconf>0.95) (http://www.sdss.org/dr4/). From this sample, we select 260928 Main galaxies in redshift region: $0.02 \leq z \leq 0.2$.

In calculating the distance we use a cosmological model with a matter density $\Omega_0 = 0.3$, cosmological constant $\Omega_A = 0.7$, Hubble's constant $H_0 = 100 h \text{km} \cdot \text{s}^{-1} \cdot \text{Mpc}^{-1}$ with h=0.7.

## 3. Selection criteria and the CG catalog

The best known catalog of compact groups is that of the Hickson Compact Groups (HCGs; Hickson 1982). From the red (E) prints of the Palomar Observatory Sky Survey (POSS), Hickson extracted 100 CGs using the following criteria:

- $N \geq 4$ (population),

- $\theta_N \geq 3 \times \theta_G$ (isolation), and

- $\mu_G < 26.0$ mag arcsec$^{-2}$ (compactness),

where
- $N$ is the total number of galaxies within 3 magnitudes of the brightest group member,

- $\mu_G$ is the total magnitude of these galaxies per square arcsec averaged over the smallest circle. containing their geometric centers,

- $\theta_G$ is the angular diameter of this smallest circle, and

- $\theta_N$ is the angular diameter of the largest concentric circle that contains no other (external)

Due to its relatively unbiased qualitative criteria, this catalog has been a popular sample for studies of compact groups. Using different criteria and galaxy catalogs, many other catalogs of compact groups also have been compiled and studied (Prandoni et al. 1994; Barton et al. 1996; Allam & Tucker 2000; Merch´an, Maia & Lambas 2000; Focardi & Kelm 2002; Ramella et al. 2002; Iovino 2002; Merch´an & Zandivarez 2002; Iovino et al. 2003; Eke et al. 2004; Lee et al. 2004; Merch´an & Zandivarez 2005).

Because Hickson's criteria only considered the galaxy angular distribution, CG catalogs identified by such criteria were actually two-dimensional samples in which CGs are seriously contaminated by background/foreground galaxies. When we try to extract CGs from a catalog of galaxies with redshift (three-dimensional galaxy sample), such criteria are not quite appropriate. Barton et al. (1996) used a different version of the friends-of-friends algorithm, and compiled a catalog of 89 redshift-selected compact groups (RSCGs) in a complete magnitude-limited redshift survey. Galaxies having projected separations $\Delta D \leq 50h^{-1}$ kpc and line-of-sight velocity differences $\Delta V \leq 1000$ km s$^{-1}$ are connected and the sets of connected galaxies constitute the groups. Apparently, the velocity selection criterion will greatly decrease the contamination by background/foreground galaxies. Unlike Hickson, Barton et al. did not include isolation and luminosity criteria, and also did not defined the minimum number of members of CGs. They only considered the galaxy spatial distribution. Because the criterion of radial distance is far more large than that of the projected separation, this algorithm is actually the quasi-three-dimensional method.

In this paper, we use the catalog of compact groups identified from the Main galaxy sample of SDSS Data Release 4 (Adelman-McCarthy et al. 2006) by Deng et al. (2006). Like Barton et al.(1996), Deng et al. (2006) only considered the galaxy spatial distribution when identifying CGs. In such CG samples, the correlations of some properties among member galaxies of CGs may be real physical effects.

Deng et al. (2006) used the conventional three-dimensional cluster analysis (Einasto et al. 1984) by which the galaxy sample can be separated into individual systems at a given neighbourhood radius R. Starting from one galaxy of the sample, we search all galaxies within a sphere of radius R around it, and call these close galaxies "friends". These "friends" and the starting galaxy are considered belonging to the same system. Around new neighbours, we continue above procedure using the rule "any friend of my friend is my friend". When no more new neighbours or "friends" can be added, then the procedure stops and a system is identified. Apparently, at small radii, most systems are some isolated single galaxies, the rest being close double and multiple galaxies. At larger radii groups and clusters of galaxies and even superclusters will be formed. By selecting different neighbourhood radii, we can probe the structures at different scales. In Deng et al. (2006)'s work, systems with $4 \leq N < 10$ (N is the number of member galaxies in each system) were selected as candidate groups. Additionally, They also explore all systems with $\geq 4$ members.

It is important to realize that we do not have any a priori defined neighbourhood radius to identify CGs. This forces us to analyse the clustering properties of galaxy sample in a certain range of neighbourhood radii. Fig.1 shows the change of the number of candidate groups and systems with $\geq 4$ members with the growth of neighbourhood radius R. In the neighbourhood radius region: R<0.4 Mpc, most systems forming in the galaxy sample are isolated single galaxies , close double and multiple galaxies, the few being candidate groups. With the growth of neighbourhood radius R the number of candidate groups and systems with $\geq 4$ members rapidly increases. In the neighbourhood radius region: R<1.2 Mpc the number of candidate groups is close to that of systems with $\geq 4$ members. Apparently, systems with $\geq 4$ members forming in this neighbourhood radius region are mostly candidate groups. These candidate groups may be CGs in the galaxy sample. Because compact groups are often located within the bounds of loose groups and clusters (Vennik et al. 1993; Ramella et al. 1994; Rood & Struble 1994; Sakai et al. 1994; Garcia 1995; Barton et al. 1996), some CGs will merge into loose groups and clusters when neighbourhood radii continue to increase. Fig.1 shows that at the neighbourhood radius $R \approx 1.2$Mpc the difference between the number of candidate groups and that of systems with $\geq 4$ members begins to become apparent. When the radius reaches about 4.4 Mpc (in the Main galaxy sample, the minimum redshift difference between galaxies is $\Delta z = 0.001$ corresponding to the minimum radial luminosity distance $\Delta D_{rad} \approx 4.4 Mpc$), systems at different redshift begin to merge into larger systems, and the number of candidate groups and systems with $\geq 4$ members decrease sharply. Fig.2 shows that the galaxy number $N_{max}$ of the richest system (it contains the largest number of member galaxies) changes with neighbourhood radius R. As seen as this figure, galaxy number $N_{max}$ of the richest system begins to increase rapidly at the neighbourhood radius $R \approx 4.4 Mpc$.

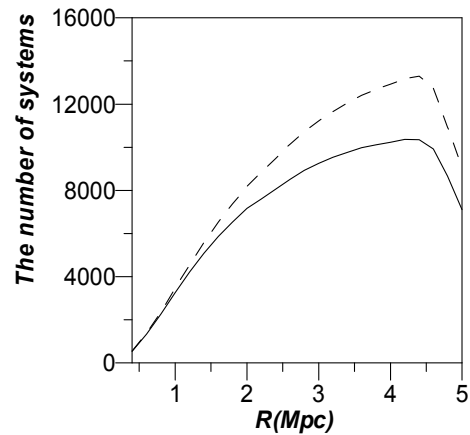

Fig.1 The change of the number of candidate groups (solid line) and systems with $\geq 4$ members (dashed line) with the growth of neighbourhood radius R.

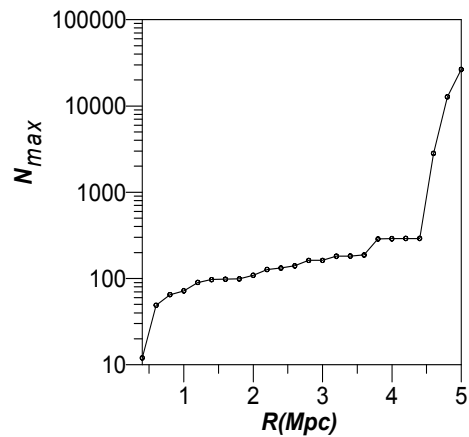

Fig.2 Relation of the galaxy number $N_{max}$ of the richest system with neighbourhood radius R.

Candidate groups identified at neighbourhood radius R=1.2 Mpc are selected as our CG sample which contains 4217 CGs (totol number of member galaxies is 21166). The selection of this neighbourhood radius mainly depends on two factors: (1) Most candidate groups do not merge into loose groups and clusters (When the neighbourhood radius is larger, many candidate groups will be included into loose groups and clusters). (2) In order to make ideal statistical analyses, we hope that our CG sample is as large as possible. As compared with previous

CG samples, this sample has two advantages: (1) Group member galaxies are located at the same redshift. So the contamination by background/foreground galaxies is completely eliminated. (2) Because our seletion criteria are only based on the galaxy spatial distribution, the correlations of some properties among member galaxies of CGs in such CG sample may be real physical effects.

### 4. The proportion of different morphological type galaxies in CGs

By removing member galaxies of CGs from the Main galaxy sample of SDSS4, we have constructed a field sample which contains 239762 Main galaxies. For each CG of the CG sample, we randomly extract a group of galaxies having the same redshift and number of galaxies as this CG from the field sample , and call them a random group. Such random group sample will reasonably sample the field with little contamination from CGs. Apparently, it has a redshift distribution completely identical to that of the original 21166 CG galaxies.

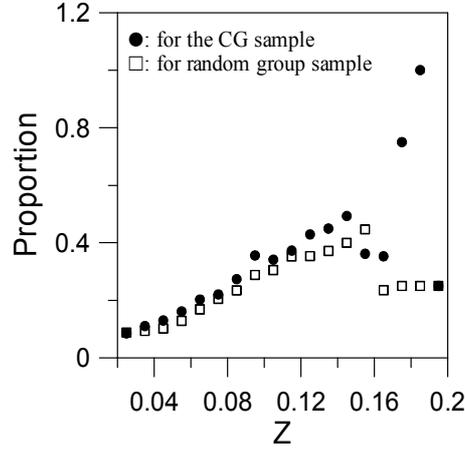

Fig.3 The proportion of early-type galaxies in different redshift bins for the CG sample and random group sample.

The concentration index $c_i = R_{90} / R_{50}$ can be used to separate early-type (E/S0) galaxies from late-type (Sa/b/c, Irr) galaxies (Shimasaku et al. 2001). Nakamura et al. (2003) confirmed that $c_i$ =2.86 separates galaxies at S0/a with a completeness of about 0.82 for both late and early types. According to Rood & Williams (1989) and Kindl (1990)'s studies, compact groups contain a significantly smaller fraction of late-type (spiral and irregular) galaxies than do their neighborhoods. Because the basic properties of Main galaxies apparently change with redshift (on the average, with growing redshift z the luminosities and sizes of galaxies, the proportion of early-type galaxies increase) (Deng et al.2005), we calculate the proportion of early-type galaxies in different redshift bins (bin $\Delta z = 0.01$) for the CG sample, and compare it with that for random group sample in the same redshift region..In Fig.3, we observe that except that in the high redshift region: $0.16 \leq z \leq 0.2$ (CGs identified in the high redshift region $0.16 \leq z \leq 0.2$ are quite few, for example, only two CGs in the redshift region $0.17 \leq z \leq 0.20$, so analysing results in this redshift region may be lack of high significance), the proportion of early-type galaxies in different redshift bins for the CG sample is statistically higher than that for random group sample, and with growing redshift z this kind of difference becomes more significant.

In Fig.3, we again notice that the proportion of early-type galaxies apparently increase with growing redshift z. This result may be due to the existence of the correlation between morphological type and luminosity. Fig.4 shows the proportion of early-type galaxies in different luminosity bins (bin $\Delta M_r = 0.4$) for the whole Main galaxy sample. The absolute magnitude $M_r$ is the r-band absolute magnitude. In our work, the K-correction (Blanton et al. 2003) is ignored. As seen from Fig.4, the proportion of early-type galaxies apparently increase with growing luminosity $M_r$-especially for bright galaxies($-22.0 \leq M_r \leq -20.0$). Because the Main galaxy sample is an apparent-magnitude limited sample (with growing redshift z the number of bright galaxies increase), the correlation mentioned above results in the proportion of early-type galaxies increasing with growing redshift z.

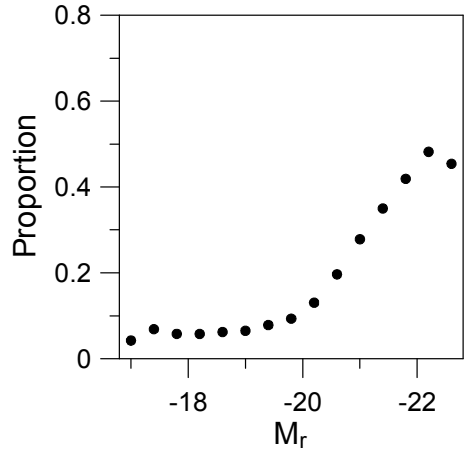

Fig.4 The proportion of early-type galaxies in different luminosity bins for the whole Main galaxy sample.

Fig.5 shows the mean luminosity of galaxies in different redshift bins for the CG sample and random group sample. We also calculate the luminosity difference $\Delta M_r$ between the brightest member and the faintest member in each CG and random group, and illustrates the distribution of this kind of the luminosity difference $\Delta M_r$ for the CG sample and random group sample in Fig.6. In order to all-roundly compare statistical properties of galaxy luminosity in the CG sample with those in random group sample, Fig.7 and Fig.8 further show the luminosity distribution of the brightest members and that of the faintest members for the CG sample and random group sample. As seen as these figure, statistical properties of galaxy luminosity in the CG sample are the same as those in random group sample. So, we can infer that the difference of the proportion of early-type galaxies between the CG sample and random group sample is not due to the correlation between morphological type and luminosity. According to Toomre (1977)'s studies, it can be explained as further evidence of interactions and mergers within a significant fraction of SDSS CGs.

The interaction often occur in a large fraction of galaxies in compact groups. The strongest direct support comes from kinematical studies. Rubin et al (1991) observed 32 HCG spiral galaxies, and found that two thirds of them have peculiar rotation curves which show characteristic of strong gravitational interaction. In their study, 12 HCG elliptical galaxies were also observed. Rubin et al (1991) detected nuclear emission in 11 of them. High fraction suggests that interactions and mergers may be supplying gas to these galaxies. By an analysis of optical images, Mendes de Oliveira & Hickson (1994) concluded that 43% of all HCG galaxies show morphological and/or kinematical distortions indicative of interaction and/or merging, and that about 32% of all HCGs contain three or more galaxies which show some sign of interaction.

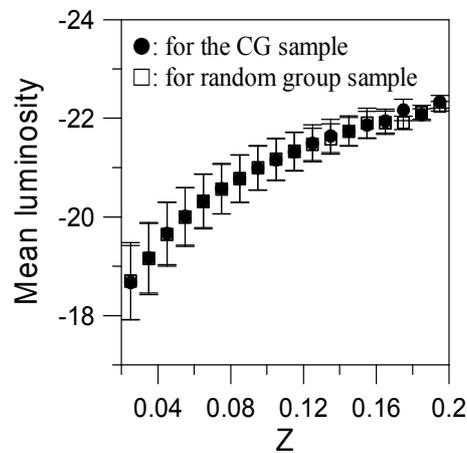

Fig.5　The mean luminosity of galaxies in different redshift bins for the CG sample and random group sample. Error bars represent standard deviation in each redshift bin.

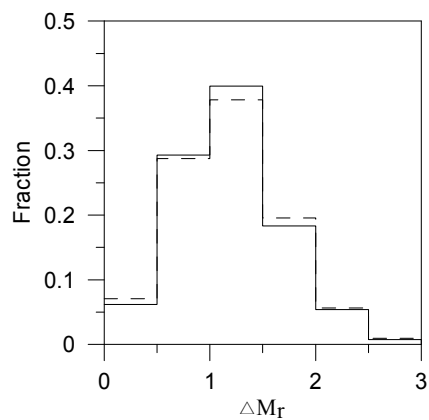

Fig.6　The distribution of the luminosity difference $\Delta M_r$ between the brightest galaxy and the faintest galaxy in each CG (solid line) and random group (dashed line).

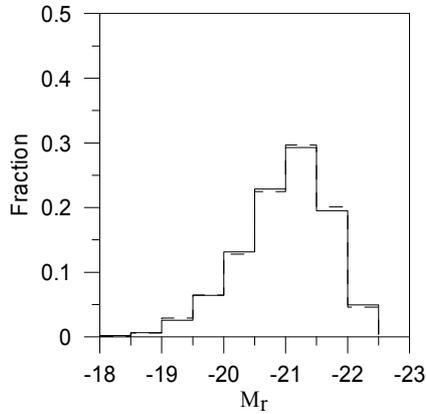

Fig.7   The luminosity distribution of the brightest galaxies for the CG sample (solid line) and random group sample(dashed line).

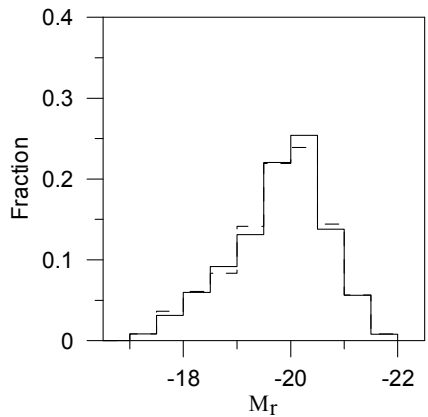

Fig.8   The luminosity distribution of the faintest galaxies for the CG sample (solid line) and random group sample (dashed line).

Hickson (1982)' analyses showed that the proportion of spiral galaxies decreases from 60% in the least compact groups to 20% in the most compact. In order to explore the the galaxy proportion of different morphological types in groups depending on compactness of groups, we construct a more compact group sample and a pair sample. The galaxy pair sample which contains 3342 galaxies are identified at radius R=100kpc (Deng et al. 2005). More compact groups are identified at radius R=0.6 Mpc. We remove galaxy pairs from  more compact group sample. Final sample of more compact groups contains 5757 galaxies. Similarly, we also remove more compact groups from the CG sample identified at radius R=1.2 Mpc. Final CG sample contains 16444 galaxies. Apparently, in pair sample the distance between members is smallest while in the CG sample that is largest. In our work, the distance between members is defined as a parameter reflecting compactness of galaxy systems.

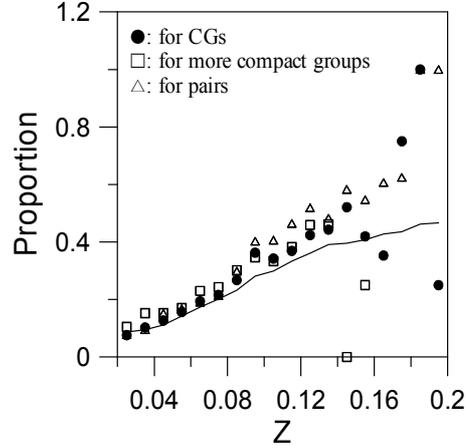

Fig.9 The proportion of early-type galaxies in different redshift bins for CGs, more compact groups and pairs. Solid line represents the proportion of early-type galaxies in different redshift bins for all Main galaxies.

Fig.9 illustrates the proportion of early-type galaxies in different redshift bins for CGs, more compact groups and pairs. Solid line shows the proportion of early-type galaxies in different redshift bins for all Main galaxies. In high redshift region $0.10 \leq z \leq 0.14$, pair sample has higher proportion of early-type galaxies than CGs and more compact groups. We infer that it can result from higher proportion of interactions and mergers in pairs than in other galaxy systems. But we do not observe as large statistical difference as Hickson (1982)' results for systems having different compactness. Additionally, above three samples having different compactness have apparently higher proportion of early-type galaxies than the whole Main galaxy sample-especially in higher redshift region. This indicates the existence of interactions and mergers within these compact galaxy systems.

## 5. Summary

In our work, the galaxy proportion of different morphological types in groups depending on compactness of groups is explored. From the field galaxy sample, we have constructed a random group sample which have the same distributions of redshift and number of member galaxies as those of the CG sample, and find that the proportion of early-type galaxies in different redshift bins for the CG sample is statistically higher than that for random group sample, and with growing redshift z this kind of difference becomes more significant. Further studies show this property is not due to the correlation between morphological type and luminosity. It can be explained as further evidence of interactions and mergers within SDSS CGs.

We also compare statistical results of CGs with those of more compact groups and pairs, but do not observe as large statistical difference as Hickson (1982)' results. Fig.9 also shows these galaxy systems having different compactness have apparently higher proportion of early-type galaxies than the whole Main galaxy sample-especially in higher redshift region.

This may indicate the existence of interactions and mergers within these compact galaxy systems.

**Acknowledgements**

Our study is supported by the National Natural Science Foundation of China (10465003).

Funding for the creation and distribution of the SDSS Archive has been provided by the Alfred P. Sloan Foundation, the Participating Institutions, the National Aeronautics and Space Administration, the National Science Foundation, the U.S. Department of Energy, the Japanese Monbukagakusho, and the Max Planck Society. The SDSS Web site is http://www.sdss.org/.

The SDSS is managed by the Astrophysical Research Consortium (ARC) for the Participating Institutions. The Participating Institutions are The University of Chicago, Fermilab, the Institute for Advanced Study, the Japan Participation Group, The Johns Hopkins University, Los Alamos National Laboratory, the Max-Planck-Institute for Astronomy (MPIA), the Max-Planck-Institute for Astrophysics (MPA), New Mexico State University, University of Pittsburgh, Princeton University, the United States Naval Observatory, and the University of Washington.

**References**

Adelman-McCarthy J.K., Agüeros M. A., Allam S.S. et al., 2006, ApJS, 162, 38
Allam S., Tucker D., 2000, AN, 321, 101
Barton E., Geller M.J., Ramella M. et al., 1996, AJ, 112, 871
Deng X.F., Jiang P., Wu P. et al., 2005, AstL, submitted
Deng X.F., Luo C.H., Jiang P. et al. 2006, APh, submitted
Einasto, J., Klypin, A. A., Saar, E., et al., 1984, MNRAS, 206, 529
Eisenstein D.J., Annis J., Gunn J. E. et al., 2001, AJ, 122, 2267
Eke V.R., Baugh C. M., Cole S.et al., 2004, MNRAS, 348, 866
Focardi P., Kelm B., 2002, A&A, 391, 35
Fukugita M., Ichikawa T., Gunn J. E. et al., 1996, AJ, 111, 1748
Garcia A., 1995, A&A, 297, 56
Gisler G., 1980, AJ, 85, 623
Gunn J. E., Carr M. A., Rockosi C. M. et al., 1998, AJ, 116, 3040
Hickson P., 1982, ApJ, 255, 382
Hickson P., Kindl E., Huchra J.P., 1988, ApJ, 331, 64
Hickson, P., 1997, ARA&A, 35, 357
Hogg D. W., Finkbeiner D. P., Schlegel D. J. et al., 2001, AJ, 122, 2129
Iovino A., 2002, AJ, 124, 2471
Iovino A., de Carvalho R. R., Gal R. R. et al., 2003, AJ, 125, 1660
Kindl E., 1990, A photometric and morphological study of compact groups of galaxies and their environments, Ph.D. Thesis, University of British Columbia,Vancouver, Canada, 168 pp
Lee B.C., Allam S.S., Tucker D.L. et al., 2004, AJ, 127, 1811

Mendes De Oliveira C., Hickson P., 1994, ApJ, 427, 684

Merch´an M.E., Maia M.A.G., Lambas D.G., 2000, ApJ, 545, 26

Merch´an M.E., Zandivarez A., 2002, MNRAS, 335, 216

Merch´an M.E., Zandivarez A., 2005, APJ, 630, 759

Nakamura O., Fukugita M., Yasuda N. et al., 2003, AJ, 125, 1682

Nilson P.N., 1973, Uppsala General Catalogue of Galaxies, Uppsala Obs. Ann. 6

Petrosian V., 1976, ApJ, 209, L1

Pier J. R., Munn J. A., Hindsley R. B.et al., 2003, AJ, 125, 1559

Prandoni I., Iovino A., MacGillivray H. T., 1994, AJ, 107, 1235

Ramella M., Diaferio A., Geller M.J. et al., 1994, AJ, 107, 1623

Ramella M., Geller M.J., Pisani A., 2002, AJ, 123, 2976

Rood H.J., Williams B.A., 1989, ApJ, 339, 772

Rood H.J., Struble M.F., 1994, PASP, 106, 413

Rubin V.C., Hunter D.A., Ford W.K.J., 1991, ApJS, 76, 153

Sakai S., Giovanelli R., Wegner G., 1994, AJ, 108, 33

Shimasaku K., Fukugita M., Doi M. et al., 2001, AJ, 122, 1238

Smith J. A., Tucker D. L, Kent S. M. et al., 2002, AJ, 123, 2121

Stoughton C., Lupton R. H., Bernardi M. et al., 2002, AJ, 123, 485

Strateva I., et al., 2001, AJ, 122, 1861

Strauss M. A., Weinberg D. H., Lupton R. H. et al., 2002, AJ, 124, 1810

Sulentic J.W., 1987, ApJ, 322, 605

Toomre A., 1977, in Evolution of Galaxies and Stellar Populations, eds. B. M. Tinsley & R. B.Larson, (New Haven: Yale University Observatory), p. 401

Vennik J., Richter G.M., Longo G., 1993, AN, 314, 393

Williams B.A., Rood H.J., 1987, ApJS, 63, 265

York D. G., Adelman J., Anderson J. E. et al., 2000, AJ, 120, 1579